\documentclass{PoS}

\usepackage{amssymb}
\usepackage{graphicx}
\usepackage{epsfig}

\newcommand{\re}[1]{Eq.~(\ref{#1})}

\newcommand{\ds}{\displaystyle}

\newcommand{\hsp}{\hspace*{1pt}}

\title{Hydrodynamic Flow and Jet Induced Mach Shocks at RHIC and LHC}

\ShortTitle{Hydrodynamic Flow and Jet Induced Mach Shocks at RHIC and LHC}

\author{\speaker{Horst~St\"ocker}\\
        FIAS- Frankfurt Institute for Advanced Studies,\\
        Max-von-Laue Str.~1, 60438 Frankfurt, Germany,\\
        Institut f\"ur Theoretische Physik, Johann Wolfgang
        Goethe - Universit\"at,\\
        Max-von-Laue Str.~1, 60438 Frankfurt, Germany\\
        E-mail: \email{stoecker@fias.uni-frankfurt.de}}

\author{Barbara~Betz\\
        Institut f\"ur Theoretische Physik, Johann Wolfgang
        Goethe - Universit\"at,\\
        Max-von-Laue Str.~1, 60438 Frankfurt, Germany\\
        E-mail: \email{betz@th.physik.uni-frankfurt.de}}

\author{Philip~Rau\\
        Institut f\"ur Theoretische Physik, Johann Wolfgang
        Goethe - Universit\"at,\\
        Max-von-Laue Str.~1, 60438 Frankfurt, Germany\\
        E-mail: \email{rau@th.physik.uni-frankfurt.de}}

\abstract{We discuss the present collective flow signals for the phase
  transition to quark-gluon plasma (QGP) and the collective flow as a
  barometer for the equation of state (EoS). A study of Mach shocks
  induced by fast partonic jets propagating through the QGP is
  given. We predict a significant deformation of Mach shocks in
  central Au+Au collisions at RHIC and LHC energies as compared to the
  case of jet propagation in a static medium. Results of a
  hydrodynamical study of jet energy loss are presented.}

\FullConference{The 3rd edition of the International Workshop --- The
  Critical Point and Onset of Deconfinement --- \\
		 July 3-7 2006\\
		 Galileo Galilei Institute, Florence, Italy}

\begin{document}

\section{The QGP phase transition}

Lattice QCD calculations yield a phase
diagram~\cite{Fodor04,Karsch04} (Fig.~\ref{phasedia}) showing a
crossing, but no first-order phase transition to the  quark-gluon
plasma (QGP) for vanishing or small chemical potentials $\mu_B$,
i.e. for conditions accessible at central rapidities at full RHIC
energy. A first-order phase transition is expected to occur only at
high baryochemical potentials or densities, i.e. at lower SPS and
RHIC energies ($\sqrt{s}\approx 4-12$ A GeV) and in the fragmentation
region of RHIC, $y \approx 3-5$ \cite{Anishetty80,Date85}. Here, the
critical baryochemical potential is predicted~\cite{Fodor04,Karsch04}
to be large, $\mu_B^c \approx 400 \pm 50 \mbox{ MeV}$, and the
critical temperature to be $T_c \approx 150-160$ MeV. We expect a
first-oder phase transition also at finite
strangeness~\cite{Greiner:1987tg}. Predictions for the phase diagram of
strongly interacting matter for realistic non-vanishing net
strangeness are urgently needed to obtain a comprehensive picture of
the QCD phase structure in all relevant dimensions (isospin,
strangeness, non-equilibrium) of the EoS. Multi-strange degrees of
freedom are very promising probes for the properties of the dense and
hot matter~\cite{Koch86}.

\subsection{Thermodynamics in the $T$- $\mu_B$ plane}

\begin{figure}[t]
\centerline{\epsfig{file=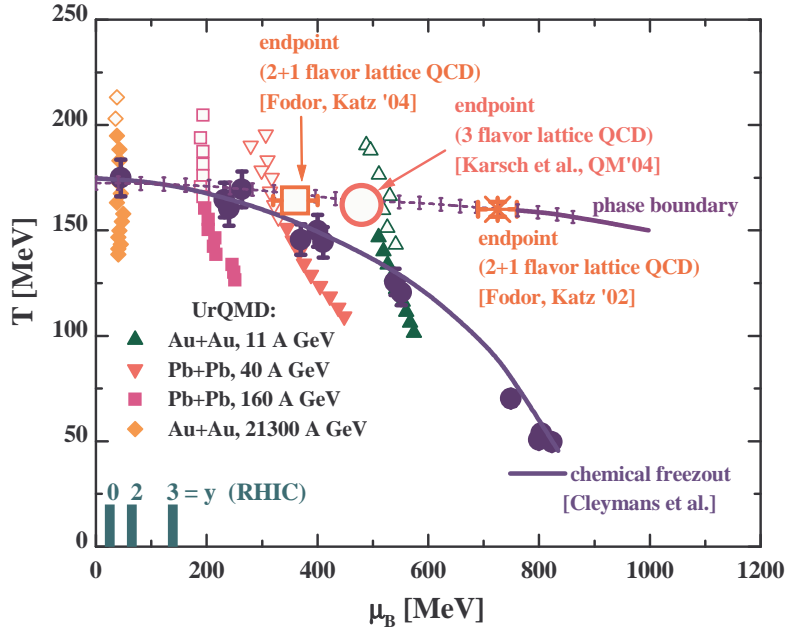,scale=0.55}}
\caption[]{Phase diagram with the critical end point at $\mu_B
  \approx 400 \mbox{ MeV}, T \approx 160 \mbox{ MeV} $ as predicted by
  Lattice QCD calculations. In addition, the time evolution in the
  $T-\mu_B$--plane of a central cell in UrQMD calculations (from
  Bravina {\it et al.})~\cite{Bravina} is depicted for different
  bombarding energies. Note that the calculations indicate that
  bombarding energies $E_{Lab} \lesssim 40$ A GeV are needed to
  probe a first-order phase transition. At RHIC this point is
  accessible in the fragmentation region only (from Bratkovskaya {\it
    et al.})\protect{~\cite{Bratkov04}}.}
\label{phasedia}
\end{figure}

Figure~\ref{phasedia} shows a comparison of the QCD predictions with
the thermodynamic parameters $T$ and $\mu_B$ extracted
from the UrQMD transport model in the central overlap
regime of Au+Au collisions~\cite{Bratkov04}. Full dots with errorbars
denote the  'experimental' chemical freeze-out parameters --
determined from fits to the experimental yields -- taken from
Ref.~\cite{Cleymans}. Triangular and quadratic symbols (time-ordered in
vertical sequence) stand for temperatures $T$ and chemical potentials
$\mu_B$ taken from UrQMD transport calculations in central Au+Au
(Pb+Pb) collisions at RHIC~\cite{Bravina} as a function of the reaction
time (separated by 1 fm/c steps from top to bottom). Open symbols
denote nonequilibrium configurations and correspond to $T$ parameters
extracted from the transverse momentum distributions, whereas the full
symbols denote configurations in approximate pressure equilibrium in
longitudinal and transverse direction.

During the nonequilibrium phase (open symbols) the transport
calculations show much higher temperatures (or energy densities) than
the 'experimental' chemical freeze-out configurations at all bombarding
energies ($\geq$ 11 A GeV). These numbers are also higher than
the critical point (circle) of (2+1) flavor lattice QCD calculations
by the Bielefeld-Swansea-collaboration~\cite{Karsch04} (large open
circle) and by the Wuppertal-Budapest-collaboration~\cite{Fodor04} (the
star denotes earlier results from~\cite{Fodor04}). The energy density at
$\mu_c, T_c$ is of the order of $\approx$ 1 GeV/fm$^3$. 
At RHIC energies a cross-over is expected at midrapidity, when
the temperature drops during the expansion phase of the 'hot
fireball'. The baryon chemical potential $\mu_B$ has been obtained
from a statistical model analysis by the BRAHMS collaboration based on
measured antihadron to hadron ratios~\cite{BRAHMS_PRL03} for different
rapidity intervals at RHIC energies. At midrapidity one finds
$\mu_B\simeq 0$, whereas at forward rapidities $\mu_B$ increases up to
$\mu_B\simeq 130$~MeV at $y=3$. Thus only a forward rapidity
measurement ($y \approx 4-5)$ will allow to probe large $\mu_B$ at
RHIC. The STAR and PHENIX detectors at RHIC offer a unique opportunity
to reach higher chemical potentials and the first-order phase
transition region at midrapidity in the high-mu-RHIC-running at
$\sqrt{s}=4-12$ A GeV in the coming year. The International FAIR
Facility at GSI will be offering a fully devoted research program in
the next decade.

\subsection{Hydrodynamic flow}

Hydrodynamic flow and shock formation has been proposed
early~\cite{Hofmann74,Hofmann76} as the key mechanism for the creation
of hot and dense matter in relativistic heavy-ion
collisions~\cite{Lacey}. The full three-dimensional hydrodynamical flow
problem is much more complicated than the one-dimensional Landau
model~\cite{Landau}: the 3-dimensional compression and expansion
dynamics yields complex triple differential cross-sections which
provide quite accurate spectroscopic handles on the EoS. The
bounce-off, $v_1(p_T)$ (i.e., the strength of the directed flow in the
reaction plane), the squeeze-out, $v_2(p_T)$ (the strength of the
second moment of the azimuthal particle emission
distribution)~\cite{Hofmann74,Hofmann76,Stocker79,Stocker80,Stocker81,Stocker82,Stocker86},
and the
antiflow~\cite{Stocker79,Stocker80,Stocker81,Stocker82,Stocker86}
(third flow component~\cite{Csernai99,Csernai04}) serve as differential
barometers for the properties of compressed, dense matter from SIS to
RHIC. In particular, it has been
shown~\cite{Hofmann76,Stocker79,Stocker80,Stocker81,Stocker82,Stocker86}
that the disappearance or so-called collapse of flow is a direct result of a
first-order phase transition.

Several hydrodynamic models~\cite{Rischke:1995pe} have been used in the
past, starting with the one-fluid ideal hydrodynamic approach. It is
well known that this model predicts far too large flow
effects. To obtain a better description of the dynamics, viscous fluid
models have been developed~\cite{Schmidt93,Muronga01,Muronga03}. In
parallel, so-called three-fluid models, which distinguish between
projectile, target and the fireball fluid, have been
considered~\cite{Brachmann97}. Here viscosity effects appear only
between the different fluids, but not inside the individual
fluids. The aim is to have at our disposal a reliable,
three-dimensional, relativistic three-fluid model including
viscosity~\cite{Muronga01,Muronga03}.

Flow can be described very elegantly in hydrodynamics. However, also
consider microscopic multicomponent (pre-)hadron transport theory,
e.g. models like qMD~\cite{Hofmann99}, IQMD~\cite{Hartnack89}, UrQMD
~\cite{Bass98}, or HSD~\cite{Cassing99}, as control models for viscous
hydrodynamics and as background models to subtract interesting
non-hadronic effects from data.

\subsection{AGS and SPS results - a review}


Microscopic (pre-)hadronic transport models describe the formation and
distributions of many hadronic particles at AGS and SPS quite
well~\cite{Weber02}. Furthermore, the nuclear EoS has
been extracted by comparing the calculation results to flow data which
are described reasonably well up to AGS
energies~\cite{Csernai99,Andronic03,Andronic01,Soff99,Sahu1,Sahu2}.
On the other hand, ideal hydrodynamical calculations predict far too
much flow at these energies~\cite{Schmidt93}, what shows that viscosity
effects have to be taken into account.

In particular, ideal hydrodynamical calculations yield factors of two
higher for the sideward flow at SIS~\cite{Schmidt93} and AGS, while the
directed flow $p_x/m$ measurement of the E895 collaboration shows that
the $p$ and $\Lambda$ data are reproduced reasonably well~\cite{Soff99,Sto04}
by UrQMD calculations, due to the reasonable cross-sections,
i.e. realistic mean-free-path of the constituents in this hadronic
transport theory.

\begin{figure}[t]
\centerline{\epsfig{file=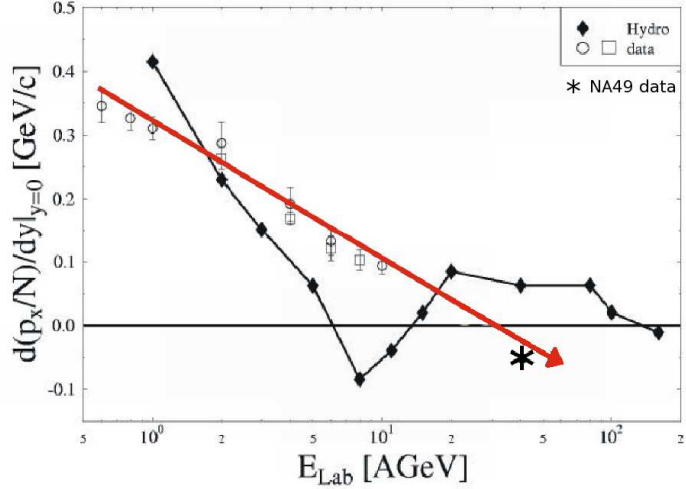,scale=0.40}}
\caption[]{Measured SIS and AGS proton $dp_x/dy$-slope data compared
  to a one-fluid hydrodynamical calculation. A linear extrapolation of
  the AGS data indicates a collapse of flow at $E_{Lab} \approx 30$ A
  GeV (see also Ref.\protect{~\cite{Brach99}}), i.e. for the lowest SPS- and 
  the upper FAIR- energies at GSI (from Paech {\it et
    al.})\protect{~\cite{Paech00}. The point at $40$~A GeV is
    calculated using the NA49 central data which clearly shows the
    proton antiflow even at near central collisions (cf. Alt {\it et
    al.})\protect{~\cite{NA49_v2pr40}}.}
\label{flow_extra}}
\end{figure}

Only ideal hydrodynamical calculations predict the appearance of a
third flow component~\cite{Csernai99} the so-called
\emph{antiflow}~\cite{Sto04,Brach00} in central collisions. We stress that
this only holds if the matter undergoes a first-order phase transition
to the QGP. The signal is that around midrapidity the directed flow,
$p_x (y)$, of protons develops a negative slope. In contrast, a
hadronic EoS without QGP phase transition does not yield such an
exotic antiflow (negative slope) wiggle in the proton flow
$v_1(y)$. The ideal hydrodynamic time evolution of the directed flow,
$p_x/N$, for the purely hadronic EoS shows a clean linear increase of
$p_x(y)$~\cite{Brach00}, just as the microscopic transport theory and as the
data~\cite{Soff99}, whereas for an EoS including a first-order phase transition
to the QGP, the proton flow  
collapses. The collapse occurs
around midrapidity. This observation is explained by an antiflow
component of protons, developing when the expansion from the
plasma sets in~\cite{Brach99}.

The ideal hydrodynamic directed proton flow $p_x$
(Fig.~\ref{flow_extra}) shows even negative values between 8 and 20 A
GeV. An increase back to positive flow is predicted with increasing
energy, when the compressed QGP phase is probed. But, where is the
predicted minimum of the proton flow in the data? Hydrodynamical
calculations suggest this ''softest-point collapse'' is at $E_{Lab}
\approx 8$ A GeV. This has not been verified by the AGS data. However,
a linear extrapolation of the AGS (Fig.~\ref{flow_extra}) data
indicates a collapse of the directed proton flow at $E_{Lab} \approx
30$ A GeV.



Recently, substantial support for this prediction has been obtained by
the low energy 40 A GeV SPS data of the NA49
collaboration~\cite{NA49_v2pr40}. These
data clearly show the first proton antiflow around mid-rapidity
(cf. Fig.~\ref{flow_extra}, in contrast to the AGS data as well as to
UrQMD calculations involving no phase transition. 

Thus, at bombarding energies of $30-40$ A GeV, the predicted effects of the
first-order phase transition to the baryon-rich QGP are most likely
observed, hence the first-order phase transition line in the
$T$-$\mu_B$-diagram has been crossed. In this energy region the
new FAIR facility at GSI will operate. There are good prospects that
the baryon flow collapse and other first-order QGP phase transition
signals can be studied soon at the lowest SPS energies as well as at
the RHIC planned HiMu-run ($\sqrt{s} = 4-12$ A GeV) at midrapidity
and possibly in the fragmentation region $y > 4-5$ for the highest
RHIC and LHC-collider energies. These experiments will enable a
detailed study of the first-order phase transition at high $\mu_B$ and
of the properties of the baryon-rich QGP in the near future.

\section{Proton elliptic flow collapse at 40 A GeV - more evidence for
  a first-order phase transition at highest net baryon densities}


At SIS energies, microscopic transport models reproduce the data on
the excitation function of the proton elliptic flow $v_2$ quite well:
A soft, momentum-dependent EoS~\cite{Andronic00,Andronic99}
seems to account for the data. The observed proton flow $v_2$ below
$\sim$ 5 A GeV is smaller than zero, which corresponds to the
squeeze-out predicted by hydrodynamics long
ago~\cite{Hofmann74,Hofmann76,Stocker79,Stocker80,Stocker81,Stocker82,Stocker86}. The
AGS data exhibit a transition from squeeze-out to in-plane flow in the
midrapidity region. The change in sign of the proton $v_2$ at $4-5$ A
GeV is in accord with transport calculations (UrQMD
calculations~\cite{Soff99} for HSD results see
Ref.~\cite{Sahu1,Sahu2}). At higher energies ($10-160$ A GeV) a smooth
increase of the flow $v_2$ is predicted from hadronic transport
simulations. In fact, the 158 A GeV data of the NA49 collaboration
suggest that this smooth increase proceeds between AGS and SPS as
predicted. Accordingly, UrQMD calculations without phase transition
give considerable 3\% $v_2$ flow for midcentral and peripheral protons
at 40 A GeV (cf. Ref.~\cite{Sto04,Soff99}).

This is in strong contrast to recent NA49 data at 40 A GeV (see
Ref.~\cite{NA49_v2pr40}): A sudden collapse of the proton flow is
observed for midcentral as well as for peripheral protons. This
collapse of $v_2$ for protons around midrapidity at 40 A GeV is
very pronounced while it is not observed at 158 A GeV.

The dramatic collapse of the flow $v_1$ also observed by
NA49~\cite{NA49_v2pr40}, again around 40 A GeV, where the collapse of
$v_2$ has been observed yields again evidence for the hypothesis
of the observation of a first-order phase transition to QCD. This is,
according to Ref.~\cite{Fodor04,Karsch04} and Fig.~\ref{phasedia}, the
highest energy at which a first-order phase transition can be reached
at central rapidities of relativistic heavy-ion collisions. We
therefore conclude that a first-order phase transition at the highest
baryon densities accessible in nature has been seen at these energies
in Pb+Pb collisions. Moreover, Ref.~\cite{Paech03} shows that the
elliptic flow clearly distinguishes between a first-order phase
transition and a crossover.

\section{ Partonic jet induced Mach shocks in an expanding QGP}
\label{sec1}

Sideward peaks have been recently
observed~\cite{StarAngCorr,Adl03b,Wan04,Jac05} in azimuthal
distributions of secondaries associated with the high-$p_T$ hadrons in
central Au+Au collisions at \mbox{$\sqrt{s}=200$\,GeV}. In
Ref.~\cite{Sto04} such peaks had been predicted as a signature of Mach
shocks created by partonic jets propagating through a QGP formed in
heavy--ion collisions. Analogous Mach shock waves were studied
previously in cold hadronic
matter\mbox{~\cite{Hofmann74,Stocker79,Stocker86,Rischke90,Cha86}} as
well as in nuclear Fermi liquids~\cite{Gla59,Kho80}. Recently, Mach
shocks from jets in the QGP have been studied in Ref.~\cite{Cas04} by
using a linearized fluid--dynamical approach.

It is well known~\cite{Landau} that a point--like perturbation moving
with supersonic speed in the spatially homogeneous ideal fluid
produces the so--called Mach region of the perturbed matter. In the
fluid rest frame (FRF) the Mach region has a conical shape with an
opening angle with respect to the direction of particle propagation
given by the expression\footnote{Quantities in the FRF are marked by tilde.} 
$
\widetilde{\theta}_M=\sin^{-1}\left(c_s/\widetilde{v}\right)\,,
$
where $c_s$ denotes the sound velocity of the unperturbed (upstream)
fluid and $\widetilde{\bf{v}}$ is the particle velocity with respect to the
fluid. In the FRF, trajectories of fluid elements (perpendicular to the
surface of the Mach cone) are inclined at the angle
$\Delta\theta=\pi/2-\widetilde{\theta}_M$ with respect to
$\widetilde{\bf{v}}$\,. Strictly speaking, the above formula is
applicable only for weak, sound--like perturbations. It is not valid
for space--time regions close to a leading particle. Nevertheless, we
shall use this simple expression for a qualitative analysis of flow
effects~\cite{Sto04,Satarov}. Following
Refs.~\cite{Sto04,Cas04,Satarov}, one can estimate the angle of
preferential emission of secondaries associated with a fast jet in the
QGP. Assuming the particle velocity to be $\widetilde{v}=1$ and the
sound velocity to be $c_s=1/\sqrt{3}$ leads to $\Delta\theta\simeq
0.96$\,. This agrees well with positions of maxima of the away--side
two--particle distributions observed in central Au+Au collisions at
RHIC energies.

\section{ Deformation of Mach shocks due to radial flow}
\label{sec2}



\begin{figure}[t]
\begin{center}
\vspace*{-5.3cm}
\epsfig{file=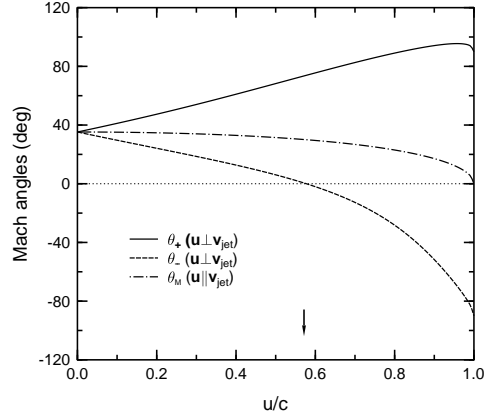,width=0.50\textwidth}
\caption[]{Angles of Mach region created by a jet moving transversely
  (solid and dashed curves) and collinearly (dashed--dotted line) to
  the fluid velocity~$\bf{u}$\, in the CMF. All curves correspond to
  $c_s^2=1/3$. The arrow marks the value $u=c_s$ (from Satarov et
  al.)\protect{~\cite{Satarov}}.}
\label{fig3}
\end{center}
\end{figure}

Assuming that the away--side jet propagates with
velocity $\bf{v}$ parallel to the matter flow velocity $\bf{u}$\, and
$\bf{u}$ does not change with space and time, one sees that after
performing the Lorentz boost to the FRF, a weak Mach
shock has a conical shape with the axis along $\bf{v}$\,. In this
reference frame, the shock front angle $\widetilde{\theta}_M$ is again given
by $\widetilde{\theta}_M=\sin^{-1}\left(c_s/\widetilde{v}\right)\,$.  
Transformation from the FRF to the center of mass frame (CMF)
shows that the Mach region remains conical, but the Mach angle becomes
smaller in the CMF, 
$\tan{\theta_M}=\left(1/\gamma_u\right)\tan{\widetilde{\theta}_M}\,,$
where $\gamma_u\equiv (1-u^2)^{-1/2}$ is the Lorentz factor
corresponding to the flow velocity~$\textbf{u}$\,.
Using the above Eqns.\ leads to the expression
for the Mach angle in the CMF
\begin{equation}\label{macp2}
\theta_M=\tan^{-1}
\left(c_s\sqrt{\frac{1-u^2}{\widetilde{v}^{\hsp 2}-c_s^2}}\right)\,,
\end{equation}
where $\widetilde{v}=(v\mp u)/(1\mp v\hsp u)$ and the upper (lower)
sign corresponds to the jet's motion in (or opposite to) the direction
of collective flow. For ultrarelativistic jets ($v\to 1$) it is
$\widetilde{v}\simeq 1$ what leads to
\begin{equation}\label{macp3}
\theta_M\simeq\tan^{-1}\left(\frac{\ds c_s\gamma_s}{\ds \gamma_u}\right)=
\sin^{-1}\left(c_s\sqrt{\frac{1-u^2}{1-u^2\hsp c_s^2}}\right)\,,
\end{equation}
with $\gamma_s=(1-c_s^2)^{-1/2}$\,. According to \re{macp3},
in the ultrarelativistic limit $\theta_M$ does not depend on the 
direction of flow with respect to the jet. The Mach cone 
becomes more narrow as compared to jet propagation in static matter. This
narrowing effect has a purely relativistic origin. Indeed, the
difference between~$\theta_M$ from \re{macp3} and the Mach angle in
absence of flow given by $\lim_{u\to 0}{\theta_M}=\sin^{-1}{c_s}$ is of
second order in the collective velocity $u$.

The case of a jet propagating at nonzero angle with respect to the flow
velocity is more complicated. Mach shocks
become nonconical for non--collinear flows. For simplicity, we
study only the case when the jet and flow velocities are
orthogonal to each other, $\bf{v}\perp\bf{u}$. Let axes $OX$ and $OY$
be directed along $\bf{u}$ and $\bf{v}$\,, respectively. We
first make the transition to the FRF by performing a Lorentz boost along
the $OX$ axis which leads to a jet velocity $\widetilde{v}$.

Assume a jet propagating along the path
$OA=\widetilde{v}\hsp\widetilde{t}$ during the time interval
$\widetilde{t}$ in the FRF. At the same time, the wave front from a
point--like perturbation (created at the origin $O$) reaches a
spherical surface with radius $OB=OC=c_s\widetilde{t}$. Two tangent
lines $AB$ and $AC$ border the Mach regio with the symmetry axis
$OA$. This region only exists for $\widetilde{v}>c_s$ what can be
fulfilled by $v>c_s$ or $u>c_s$. It has a conical shape with opening
angles $\widetilde{\theta}$ determined by the expressions
$\sin{\widetilde{\theta}} = OC/OA = c_s/\widetilde{v} \simeq{c_s}$.

Performing inverse transformation from FRF to CMF, it is easy to show
that the Mach region is modified in two ways. First, it is no longer
symmetrical with respect to the jet trajectory in the CMF.
The boundaries of the Mach wave have different angles,
$\theta_+\neq\theta_-$, with respect to $\bf{v}$ in this reference
frame. One can interpret this effect as a consequence of transverse
flow which acts like a {\it wind} deforming the Mach cone along the
direction $OX$. On the other hand, the angles of the Mach front with
respect to the beam axis are not changed under the
transformation to the CMF. We conclude that, due to effects of
transverse flow, the Mach region in the CMF should have a shape of a
deformed cone with an elliptic base. Figure~\ref{fig3} shows numerical
values of the Mach angles for an ultrarelativistic jet moving through
the QGP transversely or collinearly to its flow velocity. We point out
a much stronger sensitivity of the Mach angles $\theta_\pm$ to the
transverse flow velocity as compared with the collinear flow.


To discuss possible observable effects
we consider three events with different di--jet
axes with respect to the center of a 
fireball. 
In the first event, the away--side jet propagates along the
diameter of the fireball, i.e. collinearly with respect to the collective
flow. In the other cases, the di--jet axes are oriented along chords
close to the boundary of the fireball. In these events, the fluid
velocity has both transverse and collinear components with respect to
the jet axis. By this the
Mach fronts are deformed in an expanding matter. The radial
expansion of the fireball should cause a broadening of the sideward
peaks in the $\Delta\phi$--distributions of associated hadrons. Due to
the radial expansion, the peaks will acquire an additional width of
the order of $\left<\theta_+-\theta_-\right>$. Here $\theta_\pm$ are
local values of the Mach angles in individual events. The angular
brackets represent the  averaging over the jet trajectory in a given
event and over all events with different positions of di--jet
axes. Assuming that the particle emission is perpendicular to the surface
of Mach cone and taking $\left<u\right>\sim 0.4$, $c_s^2\simeq
1/3$, we estimate the angular spread of emitted hadrons in
the range $30^\circ-50^\circ$. This is comparable with the half
distance between the away--side peaks of the $\Delta\phi$ distribution
observed by the STAR and PHENIX
collaborations~\cite{StarAngCorr,Adl03b,Wan04,Jac05}. On the basis of
this analysis we conclude that in individual events the sideward
maxima should be asymmetric and more narrow than in an ensemble of
different events. Due to a stronger absorption of particles emitted
from the inner part of the shock, the outer two peaks may have different amplitudes. We think that these
effects can be observed by measuring three--particle correlations.
\begin{figure}[t]
\centerline{\psfig{file=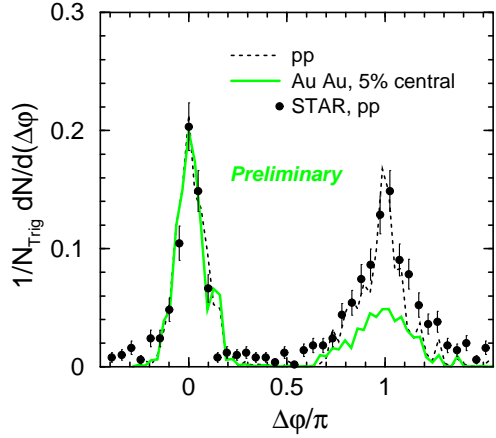,width=6cm,angle=270}}
\caption[]{STAR data on near-side and away--side jet correlation
  compared to the HSD model for p+p and central Au+Au collisions at
  midrapidity for $p_T(N_{Trig})=4\dots6\,{\rm GeV}/c$ and
  $p_T=2\,{\rm GeV}/c\dots p_T(N_{Trig})$ [from Cassing {\it et
    al.}]\protect{~\cite{Gal04,CGG}}.}
    \label{angcorr}
\end{figure}

There is one more reason for broadening of the 
$\Delta\phi$--distributions which one should keep in mind when comparing
with experimental data: due to the momentum
spread of the initial parton distributions, $\Delta p_*\lesssim 1$\,GeV,
the di--jet system has a nonzero total momentum with respect to the
global CMF. As a consequence, the angle $\theta_*$
between the trigger-- and the away--side jet is generally speaking
not equal to $\pi$\,, as was assumed above. Taking typical momenta of
initial partons as $p_0$\,, with 
$p_0>4-6$\,GeV~\cite{StarAngCorr,Adl03b,Wan04,Jac05}, we estimate the angular 
spread as $|\pi-\theta_*|\sim\Delta p_*/p_0\lesssim 0.1$\,. 
Therefore, the considered broadening should be much less than the typical
shift of the Mach angles due to the collective flow.

\section{Angular Correlations of Jets -- Can jets fake the large
$v_2$-values observed?}\label{sectionjets}
\begin{figure}[t]
\centerline{\epsfig{file=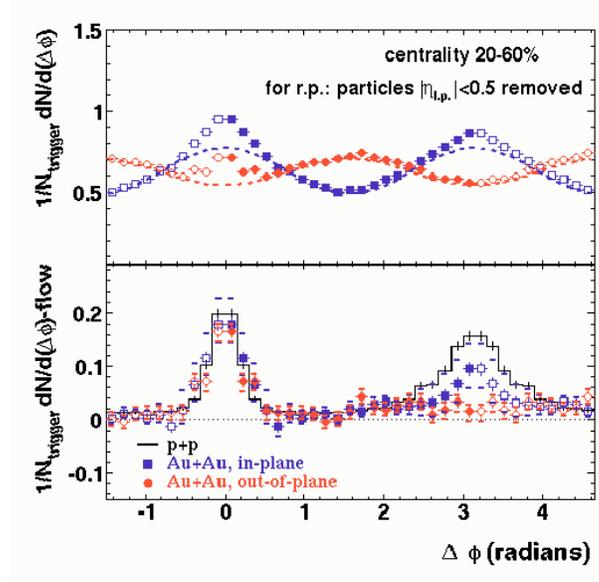,width=8cm}}
\caption[]{High $p_T$ correlations: in-plane vs. out-of-plane
  correlations of the probe (jet+secondary jet fragments) with the
  bulk ($v_2$ of the plasma at $p_T > 2\,$GeV/c) prove the existence
  of the initial plasma state (STAR-collaboration, preliminary).}
\label{filimonov}
\end{figure}

Figure~\ref{angcorr} shows 
the angular correlation of high-$p_T$ particles for the 5\% most central Au+Au
collisions at $\sqrt{s}$ = 200 GeV as well as $p+p$
reactions from the HSD-model~\cite{Gal04} in comparison to the data
from STAR for $p+p$ collisions~\cite{StarAngCorr}. Gating on
high-$p_T$ hadrons (in the vacuum) yields near--side correlations in
Au+Au collisions which are close to the near--side correlations
observed for jet fragmentation in the vacuum (p+p). This is in
agreement with the experimental
observation~\cite{StarAngCorr,Phenix}. However, for the away--side jet
correlations, the authors of Ref.~\cite {Gal04} get only a $\sim$50\%
reduction, similar to HIJING, which has only parton quenching and
neglects hadron rescattering. Clearly, the observed~\cite{StarAngCorr}
complete disappearance of the away--side jet (see
Fig.~\ref{filimonov}) cannot be explained in the HSD-(pre-)hadronic
cascade even with a small formation time of $0.8\,$fm/c. Hence, the
correlation data provide another clear proof for the existence of the
bulk plasma.

The question if the attenuation of jets of $p_T \ge5$ GeV/c can
actually fake the observed $v_2$-values at $p_T \approx 2$GeV/c comes
about since due to fragmentation and rescattering a lot of
momentum-degraded hadrons will propagate in the hemisphere defined by
the jets. However, their momentum dispersion perpendicular to the jet
direction is so large that it could indeed fake a collective flow that
is interpreted as coming from the early high-pressure plasma phase.

\begin{figure}[t]
\centerline{\epsfig{file=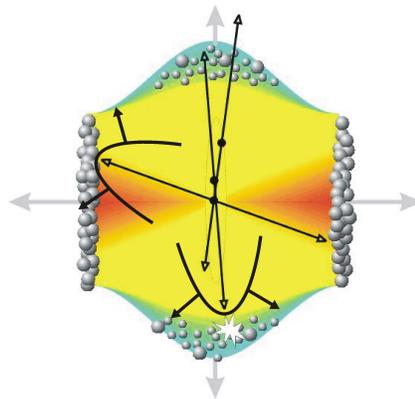,width=5.5cm}}
\caption[]{Illustration of jets traveling through the late hadronic
  stage of the reaction. Only jets from the region close to the
  initial surface can propagate and fragment in the
  vacuum\protect{~\cite{Hofmann74,Gal04,Baum75}}. The other
  jets will interact with the bulk, resulting in wakes with bow waves
  travelling transversely to the jet axis.}
\label{fig:scheme}
\end{figure}

On first sight, Fig.~\ref{filimonov} shows that this could indeed be
the case: the in-plane $v_2$ correlations are aligned with the jet
axis, the away--side bump, usually attributed to collective $v_2$ flow
(dashed line), could well be rather due to the stopped, fragmented and
rescattered away--side jet. However, this argument is falsified by the
out-of-plane correlations (circles in Fig.~\ref{filimonov}). The
near-side jet is clearly visible in the valley of the collective flow
$v_2$ distribution. Note that $v_2$ peaks at $\varphi = \pi/2$
relative to the jet axis! The away--side jet, on the other hand, has
completely vanished in the out-of-plane distribution
(cf. Fig.~\ref{fig:scheme}).

Where are all the jet fragments gone and why is there no trace left?
Even if the away--side jet fragments completely and the fragments get
stuck in the plasma, leftovers should be detected at momenta below
$2\,$GeV/c. Hadronic models as well as parton cascades will have a
hard time to get a quantitative agreement with these exciting data.

We propose future correlation measurements which can yield
spectroscopic information on the plasma:

\begin{itemize}
\item
If the plasma is a color-electric plasma~\cite{Sto04,Ruppert:2005uz}, 
experiments will - in spite of strong plasma damping - be able to
search for wake-riding potential effects. The wake of the leading jet
particle can trap comoving companions moving through the plasma in
the wake pocket with the same speed as the leading
particle. This can be particular stable for charmed jets due to the
deadcone effect (proposed by Kharzeev {\it et al.}~\cite{Kharzeev}) which
will guarantee little energy loss, i.e. constant velocity of the
leading D-meson. The leading D-meson will practically have very little
momentum degradation in the plasma and therefore the wake potential
following the D will be able to capture the equal speed companion,
which can be detected~\cite{Schafer78}.

\item
The sound velocity of the expanding plasma might be measured by the
emission pattern of the plasma particles travelling sideways with
respect to the jet axis: The dispersive wave generated by the wake of
the jet in the plasma yields preferential emission to an angle
relative to the jet axis given by the ratio of the leading
jet particles' velocity, devided by the sound velocity in the hot dense
plasma rest frame. The speed of sound for a non-interacting gas of
relativistic massless plasma particles is $c_s \approx
1/\sqrt{3} \approx 57 \% \,c$, while for a plasma with strong
vector interactions, $c_s =  c$ holds. Hence, the emission angle measurement
can yield information of the interactions in the plasma. A
hydrodynamical study of this point will be discussed in the following.
\end{itemize}

\section{(3+1)dimensional hydrodynamical study of jet evolution}

\begin{figure}[t]
\begin{center}
\begin{minipage}[t]{5.5cm}
\epsfig{file=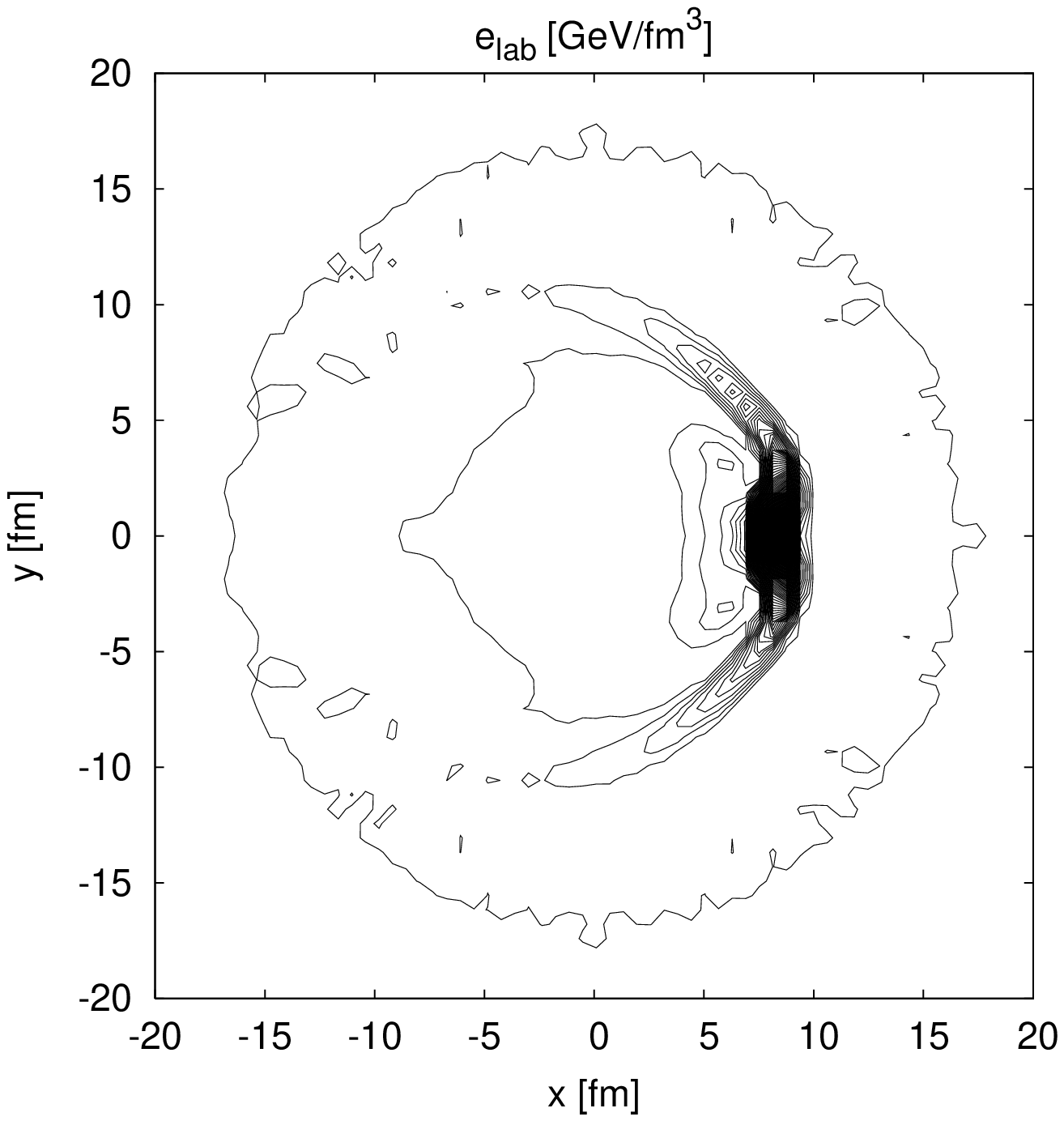,width=0.8\textwidth}
\end{minipage}
\hspace*{1cm}
\begin{minipage}[t]{5.5cm}
\epsfig{file=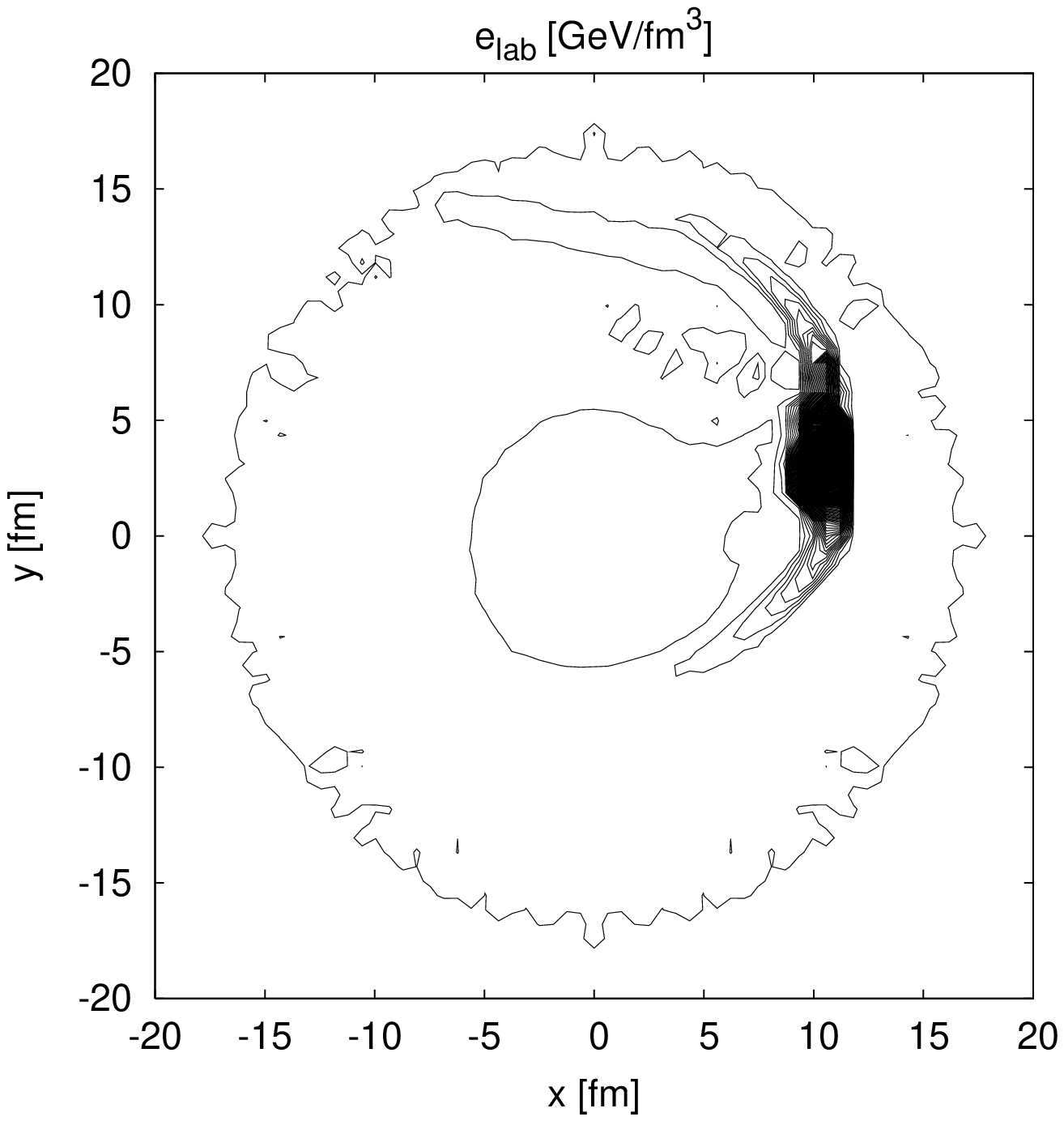,width=0.8\textwidth}
\end{minipage}\\\vspace*{0.5cm}
\begin{minipage}[t]{5.6cm}
\epsfig{file=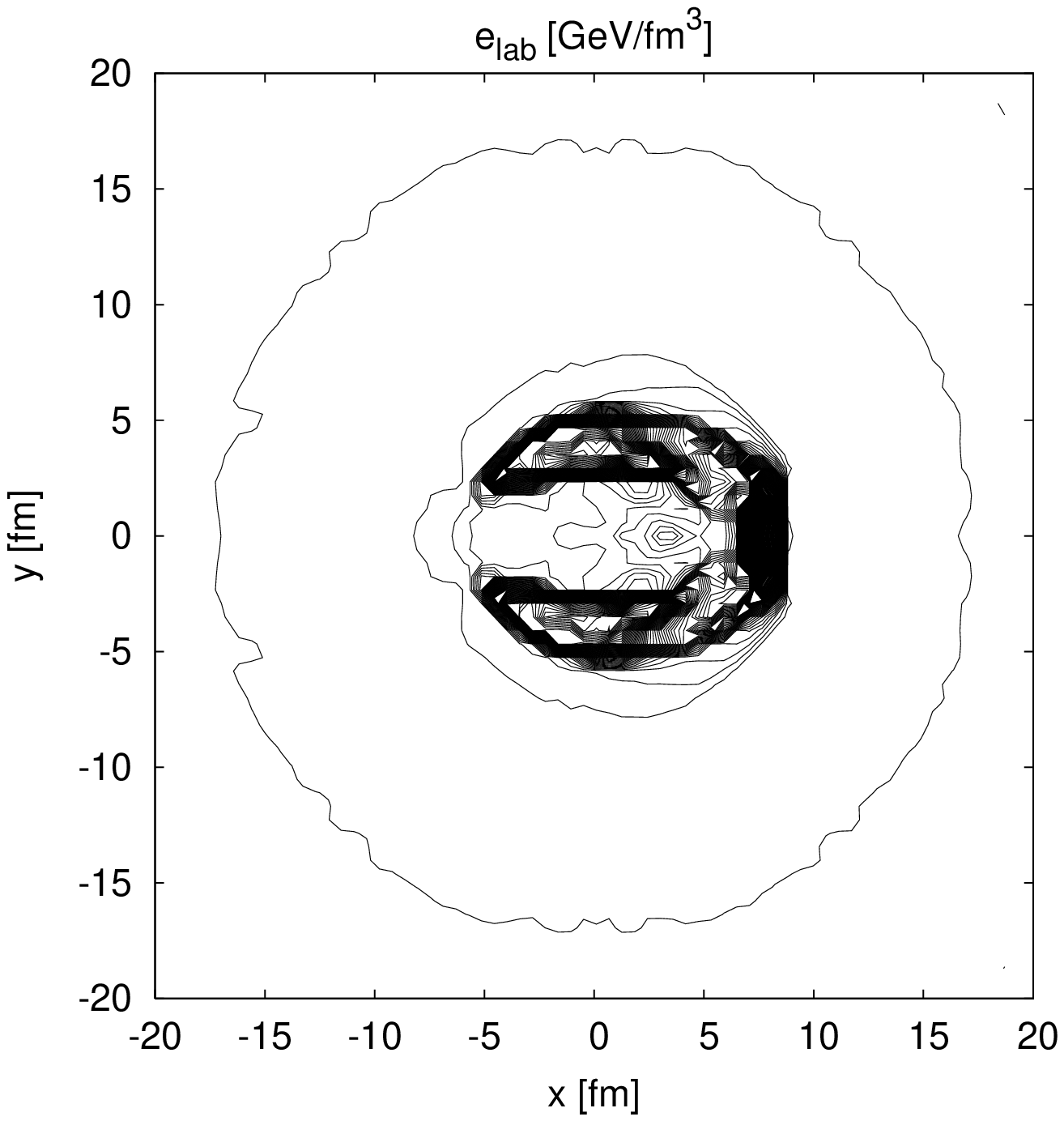,width=0.8\textwidth}
\end{minipage}
\hspace*{1cm}
\begin{minipage}[t]{5.6cm}
\epsfig{file=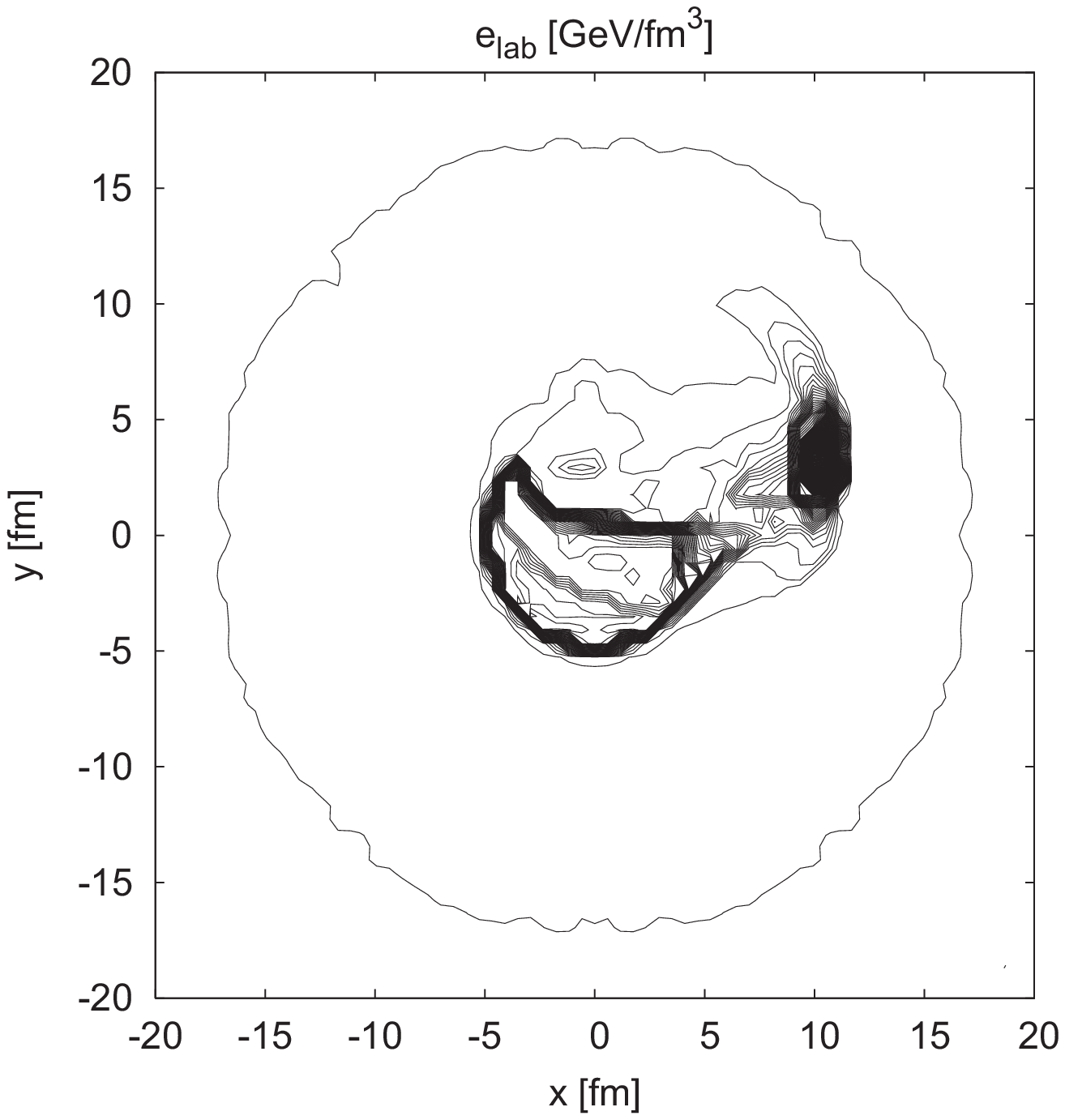,width=0.76\textwidth}
\end{minipage}
\caption{Contour plots of the laboratory energy density at $t=12.8$~fm/c
for an ideal gas EoS (upper panel) and for a hadron gas with first-order 
phase transition to QGP (lower panel) for different initial locations
of the jet (see text).\label{Jetplot}}
\end{center}
\end{figure}

The STAR and PHENIX collaborations published the
observation~\cite{StarAngCorr,STAR,Phenix} that the away--side jet in Au+Au
collisions for high-$p_T$ particles ($4 < p_T$(trigger)$< 6$~GeV/c,
$p_T$(assoc)$> 2$~GeV/c) with pseudo-rapidity $|y| < $~0.7 is
suppressed as compared to the away--side jet in p+p collisions (see
Fig.~\ref{filimonov}).

This is commonly interpreted as parton energy loss, the so-called jet
quenching~\cite{Sto04,jetquen}. One part of the back-to-back jet
created in the collision escapes (near-side jet), the other one
(away--side jet) deposits a large fraction of its energy into the dense
matter.

We use (3+1)dimensional ideal hydrodynamics, 
employ the (3+1)dimensional SHASTA (SHarp And Smooth
Transport Algorithm)~\cite{SHASTA}, and follow the time evolution of a
fake jet that deposits its energy and momentum completely during a
very short time in a 2 fm$^3$ spatial volume of a spherically
symmetric expanding system.

The medium has an initial radius of $5$~fm, an initial energy density
of $e_0=1.68\, {\rm GeV/fm}^3$ and an initial profile velocity
increasing by radius as ${\bf v}(r) =0.02\,r/R$.

The initial energy density of the jet is increased by $\Delta e=5\,
\rm{GeV/fm}^3$ as compared to the medium and the jet 
material has an initial velocity of $v_x=0.96$~c.
In Fig.~\ref{Jetplot} we display the contour plots of the jet
evolution at late state $t=12.8$~fm/c for an ultrarelativistic ideal
gas EoS (upper row) and a hadron gas with a first-order phase
transition to QGP (lower row). The jet is initially located in the
region between $-5 \,{\rm fm}<x<3\,{\rm  fm},\,|y|<0.5\,{\rm fm},\,
|z|<0.5\,{\rm fm}$ (left column) and in the retion between $-3 \,{\rm
  fm}<x<-1\,{\rm  fm},\,2.5 \,{\rm fm}<y<3.5\,{\rm  fm}, |z|<0.5\,{\rm
  fm}$ (right column).

\begin{figure}[t]
\begin{center}
\begin{minipage}[t]{5.6cm}
\epsfig{file=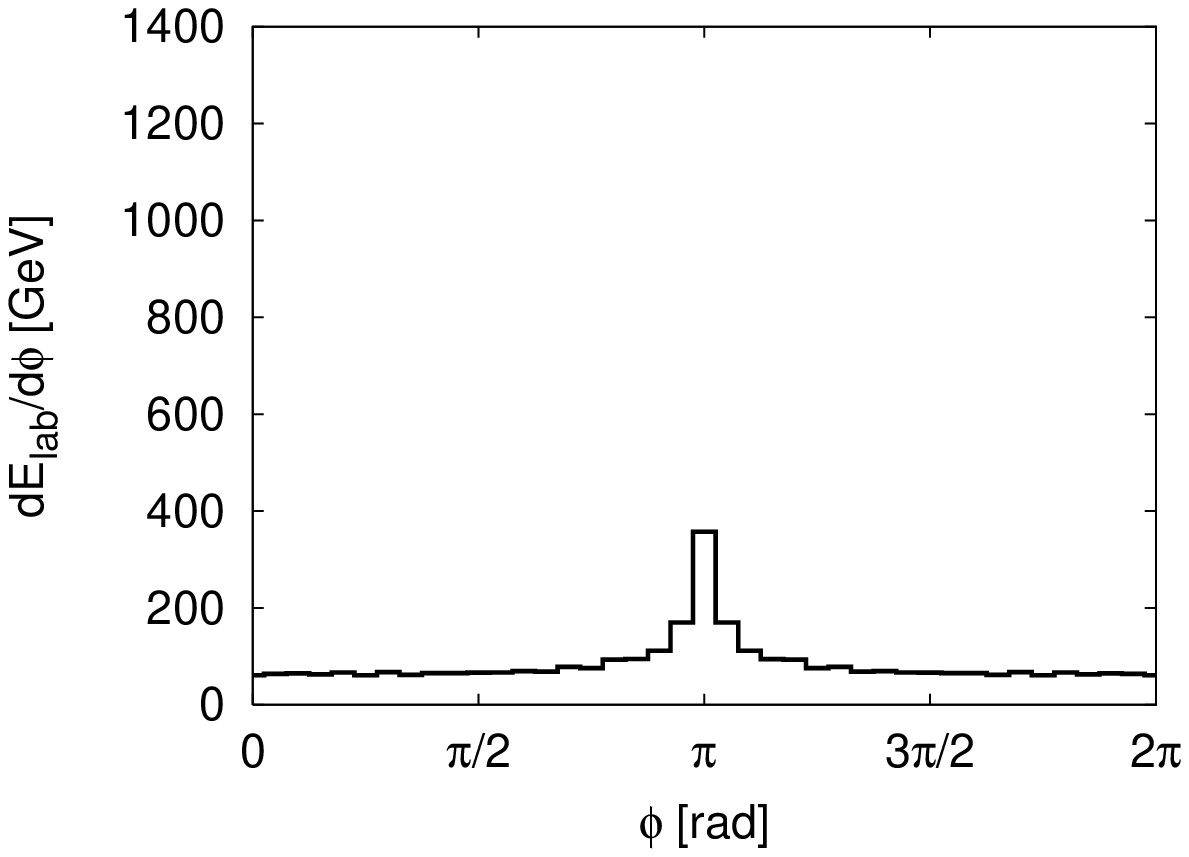,width=0.9\textwidth}
\end{minipage}
\hspace*{1cm}
\begin{minipage}[t]{5.6cm}
\epsfig{file=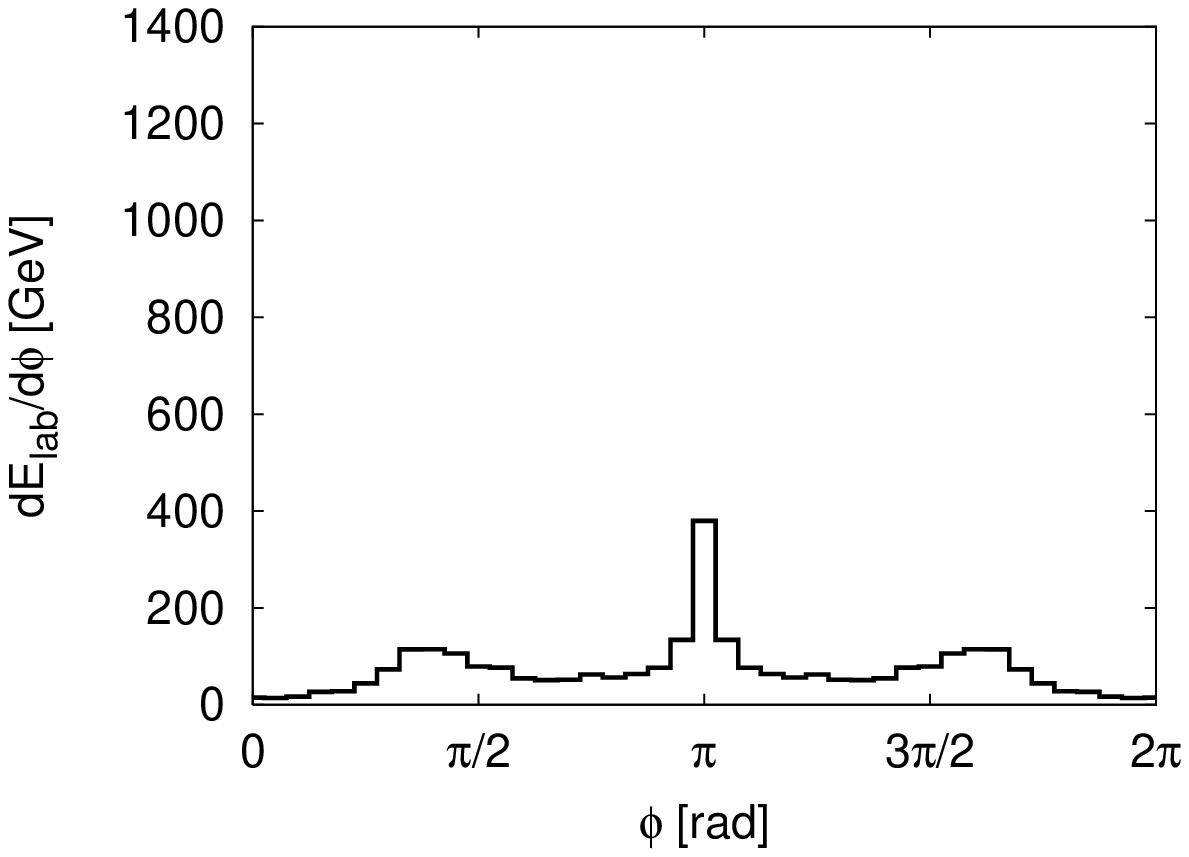,width=0.9\textwidth}
\end{minipage}
\caption{Azimuthal angular distributions of the laboratory energy at
  $t=12.8$~fm/c for an ideal gas EoS (left) and for a hadron
  gas with first-order phase transition to QGP (right). No
  background is subtrackted.\label{Distplot}}
\end{center}
\end{figure}

The jet-induced shock front and a deflection of the jet for a finite
impact parameter is clearly visible. Note, that in case of the Bag
Model EoS the system reaches the mixed-phase. Therefore, the
hydrodynamical evolution slows down and causes a broadend shock wave.

Figure~\ref{Distplot} shows the azimuthal angular distributions of the
laboratory energy at $t=12.8$~fm/c for an ideal gas EoS (left) and for a
hadron gas with first-order phase transition to QGP (right). Here, the jet was
originally located between $-5 \,{\rm fm}<x<3\,{\rm
  fm},\,|y|<0.5\,{\rm fm},\, |z|<0.5\,{\rm fm}$. One clearly sees that
sideward peaks occur in case of a first-order phase transition. They
are a signal of conical emission and agree well with the 2 and 3
particle correlations from STAR and PHENIX~\cite{StarAngCorr,STAR,Phenix}.

\section{Summary}

The NA49 collaboration has observed the collapse of both, $v_1$- and
$v_2$-collective flow of protons, in Pb+Pb collisions at 40 A GeV,
which presents evidence for a first-order phase transition in
baryon-rich dense matter. It will be possible to study the nature of
this transition and the properties of the expected chirally restored
and deconfined phase both at the HiMu/low energy and at the forward
fragmentation region at RHIC, with upgraded and/or second generation
detectors, and at the future GSI facility FAIR. According to lattice
QCD results~\cite{Fodor04,Karsch04}, the first-order phase transition
occurs for chemical potentials above 400 GeV. Ref.~\cite{Paech03} shows
that the elliptic flow clearly distinguishes between a first-order
phase transition and a crossover. Thus, the observed collapse of flow,
as predicted in Ref.~\cite{Hofmann74,Hofmann76}, is a clear signal for
a first-order phase transition at the highest baryon densities.

A critical discussion of the use of collective flow as a barometer for
the EoS of hot dense matter at RHIC showed that
hadronic rescattering models can explain $< 30 \%$ of the observed
flow, $v_2$, for $p_T > 2$ GeV/c. We interpret this as evidence for
the production of superdense matter at RHIC with initial pressure way
above hadronic pressure, $p > 1$~GeV/fm$^3$.

The fluctuations in the flow, $v_1$ and $v_2$, should be
measured. Ideal hydrodynamics predicts that they are larger than 50 \%
due to initial state fluctuations. The QGP coefficient of viscosity
may be determined experimentally from the fluctuations observed.

We propose upgrades and second-generation experiments at RHIC, which
inspect the first-order phase transition in the fragmentation region,
i.e., at $\mu_B\approx~400$~MeV ($\sqrt{s}=4-12$ A GeV or $y \approx
4-5$ at full energy), where the collapse
of the proton flow analogous to the 40 A GeV data should be seen.

The study of jet-wake-riding potentials and bow shocks caused by jets
in the QGP formed at RHIC can give further clues on the EoS and
transport coefficients of the QGP.

\section{Acknowledgements}
We like to thank M. Bleicher, I. Mishustin, K. Paech, H. Petersen,
D. Rischke and L. Satarov for discussions.



\begin{thebibliography}{99}

\bibitem{Fodor04}
    Z. Fodor and S. D. Katz,  JHEP {\bf 0203} (2002) 014;
    JHEP {\bf 0404} (2004) 050.
    
\bibitem{Karsch04}
    F. Karsch, J.\ Phys.\ G {\bf 30} (2004) S887;
    F.~Karsch,hep-ph/0701210.
    
\bibitem{Anishetty80}
     R.~Anishetty, Peter Koehler, and Larry~D. McLerran,
     Phys. Rev. {\bf D 22} (1980) 2793.
     
\bibitem{Date85}
     S.~Date, M.~Gyulassy, and H.~Sumiyoshi,
     Phys. Rev. {\bf D 32} (1985) 619.

\bibitem{Greiner:1987tg}
  C.~Greiner, P.~Koch, and H.~St\"ocker,
  Phys.\ Rev.\ Lett.\  {\bf 58} (1987) 1825;
  C.~Greiner, D.~H.~Rischke, H.~St\"ocker, and P.~Koch,
  Phys.\ Rev.\  D {\bf 38} (1988) 2797.   

 \bibitem{Koch86}
     P.~Koch, B.~M\"uller, and J.~Rafelski,
     Phys. Rept. {\bf 142} (1986) 167;
     I.~Zakout, C.~Greiner, and J.~Schaffner-Bielich,
     Nucl.\ Phys.\  A {\bf 781} (2007) 150;
     J.~Schaffner, C.~B.~Dover, A.~Gal, C.~Greiner, D.~J.~Millener, and
     H.~St\"ocker,
     Annals Phys.\  {\bf 235} (1994) 35;
     J.~Schaffner, C.~B.~Dover, A.~Gal, C.~Greiner, and H.~St\"ocker,
     Phys.\ Rev.\ Lett.\  {\bf 71} (1993) 1328.

\bibitem{Bratkov04}
    E.~L. Bratkovskaya {\it et al.},
    Phys.\ Rev.\  C {\bf 69} (2004) 054907.

\bibitem{Cleymans}
    J.~Cleymans and K.~Redlich,
     Phys. Rev. {\bf C60} (1999) 054908.

\bibitem{Bravina}
    L. V. Bravina {\it et al.},
    Phys. Rev.  {\bf C 60} (1999) 024904;
      Nucl. Phys.  {\bf A 698} (2002) 383.

\bibitem{BRAHMS_PRL03}
    I.~G. Bearden {\it et al.},
    Phys. Rev. Lett. {\bf 90} (2003) 102301.

\bibitem{Hofmann74}
    J.~Hofmann, H.~St\"ocker, W.~Scheid, and W.~Greiner,
  {Report of the Int. Workshop on BeV/Nucleon Collisions of Heavy Ions:
  How and Why}, Bear Mountain, New York, Nov. 29 - Dec. 1, 1974
  (BNL-AUI  1975).

\bibitem{Hofmann76}
    J.~Hofmann, H.~St\"ocker, U.~W. Heinz, W.~Scheid, and W.~Greiner,
    Phys. Rev. Lett. {\bf 36} (1976) 88.

\bibitem{Lacey}
  R.~A.~Lacey and A.~Taranenko, nucl-ex/0610029.

\bibitem{Landau}
    L.D. Landau and E.M. Lifshitz,
    \textit{Fluid Mechanics},
    Pergamon Press, New York, 1959.

\bibitem{Stocker79}
    H.~St\"ocker, J.~Hofmann, J.~A. Maruhn, and W.~Greiner,
    Prog. Part. Nucl. Phys. {\bf 4} (1980) 133.

\bibitem{Stocker80}
    H.~St\"ocker, J.~A. Maruhn, and W.~Greiner,
    Phys. Rev. Lett. {\bf 44} (1980) 725.

\bibitem{Stocker81}
  H.~St\"ocker {\it et al.},
  Phys. Rev. Lett. {\bf 47} (1981) 1807.

\bibitem{Stocker82}
    H.~St\"ocker {\it et al.},
    Phys. Rev. {\bf C 25} (1982) 1873.

\bibitem{Stocker86}
    H.~St{\"o}cker and W.~Greiner.
    Phys. Rept.  {\bf 137} (1986) 277.

\bibitem{Csernai99}
    L.~P. Csernai and D.~R\"ohrich,
    Phys. Lett. {\bf B 458} (1999) 454.

\bibitem{Csernai04}
    L.~P. Csernai {\it et al.},
    hep-ph/0401005.

\bibitem{Rischke:1995pe}
  D.~H.~Rischke, Y.~P\"urs\"un, J.~A.~Maruhn, H.~St\"ocker, and W.~Greiner,
  Heavy Ion Phys.\  {\bf 1} (1995) 309.

\bibitem{Schmidt93}
    W.~Schmidt {\it et al.},
     Phys. Rev. {\bf C 47} (1993) 2782.

\bibitem{Muronga01}
    A. Muronga,
    Heavy Ion Phys. {\bf 15} (2002) 337.

\bibitem{Muronga03}
    A. Muronga,
    Phys. Rev. {\bf C 69} (2004) 034903.

\bibitem{Brachmann97}
    J.~Brachmann {\it et al.},
    Nucl. Phys. {\bf A 619} (1997) 391.

\bibitem{Hofmann99}
    M.~Hofmann {\it et al.},
    nucl-th/9908031.

\bibitem{Hartnack89}
    C.~Hartnack {\it et al.},
    Nucl. Phys. {\bf A 495} (1989) 303c.

\bibitem{Bass98}
    S.~A. Bass, M.~Gyulassy, H.~St{\"o}cker, and W.~Greiner,
    J. Phys. {\bf G 25} (1999) R1.

\bibitem{Cassing99}
    W.~Cassing and E.~L. Bratkovskaya,
    Phys. Rept. {\bf 308} (1999) 65.

\bibitem{Weber02}
    H.~Weber, E.~L. Bratkovskaya, W.~Cassing, and H.~St\"ocker,
    Phys. Rev. {\bf C 67} (2003) 014904.

\bibitem{Andronic03}
    A.~Andronic {\it et al.},
    Phys. Rev. {\bf C 67} (2003) 034907.

\bibitem{Andronic01}
    A.~Andronic {\it et al.},
    Phys. Rev. {\bf C 64} (2001) 041604.

\bibitem{Soff99}
    S. Soff, S. A. Bass, M. Bleicher, H. St\"ocker, and W. Greiner,
    nucl-th/9903061.
    
\bibitem{Sahu1}
    P. K. Sahu and W. Cassing,
     Nucl. Phys. {\bf A 672} (2000) 376.

\bibitem{Sahu2}
    P.~K.~Sahu and W.~Cassing,
     Nucl. Phys. {\bf A 712} (2002) 357.
    
\bibitem{Sto04}
	H.~St\"ocker,
	Nucl. Phys. \textbf{A 750} (2005) 121. 

\bibitem{Brach00}
    J.~Brachmann,
    PhD thesis, J. W. Goethe - Universit\"at Frankfurt am Main, 2000.

\bibitem{Brach99}
    J.~Brachmann {\it et~al.},
     Phys. Rev. {\bf C 61} (2000) 024909.
     
\bibitem{Paech00}
    K.~Paech, M.~Reiter, A.~Dumitru, H.~St\"ocker, and W.~Greiner,
     Nucl. Phys. {\bf A 681} (2001) 41.

\bibitem{NA49_v2pr40}
    C.~Alt {\it et al.},
    Phys. Rev. {\bf C 68} (2003) 034903.

\bibitem{Andronic00}
    A.~Andronic {\it et al.},
     Nucl. Phys. {\bf A 679} (2001) 765.

\bibitem{Andronic99}
    A.~Andronic,
    Nucl. Phys. {\bf A 661} (1999) 333.

\bibitem{Paech03}
    K.~Paech, H.~St\"ocker, and A.~Dumitru,
    Phys. Rev. {\bf C 68} (2003) 044907;       
    Phys. Rev. {\bf C 62} (2000) 064611.

\bibitem{StarAngCorr}
    C.~Adler {\it et~al.} [STAR collaboration],
     Phys. Rev. Lett. {\bf 90} (2003) 082302;
     L.~Molnar, nucl-ex/0701061.

\bibitem{Adl03b}
	C.~Adler {\it et~al.} [STAR collaboration],
	Phys. Rev. Lett. \textbf{91} (2003) 072304;
	S.~S.~Adler {\it et al.}  [PHENIX collaboration], 
	Phys.\ Rev.\  D {\bf 74} (2006) 072002.

\bibitem{Wan04}
	Fuqiang Wang [STAR collaboration],
	J.\ Phys.\ G {\bf 30} (2004) S1299;
	J.~G.~Ulery and F.~Wang, nucl-ex/0609017;
	F.~Wang, Nucl.\ Phys.\  A {\bf 783} (2007) 157;
	F.~Wang, nucl-ex/0610027.

\bibitem{Jac05}
	B. Jacak [PHENIX collaboration], N.N. Ajitanand [PHENIX collaboration],
	talks at Int. Conf. on Physics and Astrophysics
	of Quark Gluon Plasma, Kolkata, India, 2005;
	B.~Jacak  [PHENIX collaboration], J.\ Phys.\ Conf.\ Ser.\
        {\bf 50} (2006) 22.


\bibitem{Rischke90}
    D.~H. Rischke, H.~St\"ocker, and W.~Greiner,
    Phys. Rev. {\bf D 42} (1990) 2283.

\bibitem{Cha86}
	G.F. Chapline and A. Granik,
	Nucl. Phys. \textbf{A 459} (1986) 681.

\bibitem{Gla59}
	A.E. Glassgold, W. Heckrotte, and K.M. Watson,
	Ann. Phys. \textbf{6} (1959) 1.

\bibitem{Kho80}
	V.A. Khodel, N.N. Kurilkin, and I.N. Mishustin,
	Phys. Lett. \textbf{B 90} (1980) 37.

\bibitem{Cas04}
	J. Casalderrey--Solana, E.V. Shuryak, and D. Teaney,
	J.\ Phys.\ Conf.\ Ser.\  {\bf 27} (2005) 22;
	J.~Casalderrey-Solana, hep--ph/0701257;
	F.~Antinori and E.~V.~Shuryak, J.\ Phys.\ G {\bf 31} (2005) L19.
	
\bibitem{Satarov}
  L.~M.~Satarov, H.~St\"ocker, and I.~N.~Mishustin
  Phys.\ Lett.\  B {\bf 627} (2005) 64.

\bibitem{Gal04}
	W. Cassing, K. Gallmeister, and C. Greiner,
	J. Phys. \textbf{G 30} (2004) 801.


\bibitem{CGG}
    W.~Cassing, K.~Gallmeister, and C.~Greiner,
    { Nucl. Phys.} {\bf A 735} (2004) 277.

\bibitem{Phenix}
  A.~Sickles  [PHENIX collaboration], nucl-ex/0702007.


\bibitem{Baum75}
    H.~G. Baumgardt {\it et~al.},
    Z. Phys. {\bf A 273} (1975) 359.	
    
 \bibitem{Ruppert:2005uz}
   J.~Ruppert and B.~M\"uller,
   Phys.\ Lett.\  B {\bf 618} (2005) 123.


\bibitem{Kharzeev}
    D.~Kharzeev,
    private communication.

\bibitem{Schafer78}
    W.~Sch\"afer, H.~St\"ocker, B.~M\"uller, and W.~Greiner,
    Z. Phys. {\bf A 288} (1978) 349.

\bibitem{STAR}
    J.~Adams {\it et al.}  [STAR collaboration], Phys.\ Rev.\ Lett.\
    {\bf 91} (2003) 072304.
    
\bibitem{jetquen}
    M.~Gyulassy, P.~Levai, and I.~Vitev, Nucl.\ Phys.\ B {\bf 594}
    (2001) 371.

\bibitem{SHASTA}
    D.~H.~Rischke, S.~Bernard, and J.~A.~Maruhn, 
    Nucl.\ Phys.\ A {\bf 595} (1995) 346.


\end{thebibliography}
\end{document}